\documentclass{article}
\usepackage{amssymb}

\input{tcilatex}

\begin{document}

\begin{center}
\bigskip 

\bigskip

\bigskip

{\huge BRST Exact Action for N=1, D=4 Supersymmetric Yang-Mills Theory}
\end{center}

\begin{quote}
{\huge \bigskip }
\end{quote}

\begin{center}
\bigskip

\bigskip

\bigskip

{\LARGE A. Aidaoui}$^{\ast \text{, }\intercal }${\LARGE \ and M. Tahiri}$%
^{\ast }$

{\Large \bigskip }

$^{\ast }${\Large Laboratoire de Physique Th\'{e}orique, Universit\'{e}
d'Oran Es-Senia, 31100 Oran, Algeria}

\bigskip

{\normalsize \ \ }$^{\intercal }${\Large Centre Universitaire de B\'{e}char,
B. P. 417, 08000 B\'{e}char,\ Algeria}

{\Large E-mail: hadj\_aidaoui@yahoo.fr}

{\Large \bigskip }

\bigskip

\bigskip

\bigskip

\bigskip
\end{center}

\bigskip ${\Large \baselineskip=50pt}$

${\Large \centerline{\bf Abstract}\bigskip }$

\bigskip

\begin{quote}
\ \ \ \ \ {\large \ The action for classical and consequently quantum N=1,
D=4 supersymmetric Yang-Mills theories is proven to be written as a BRST
exact term. In order to show this, the space of fields in the theory must be
enlarged to include an extrafield as a non dynamical auxiliary field}{\Large %
.}

{\Large \bigskip }

{\normalsize Keywords: Supersymmetric theories, BRST exact quantum action.}

{\normalsize \bigskip }

{\normalsize PACS numbers: 11.15.-q, 12.60.Jv}
\end{quote}

{\normalsize \bigskip }

\bigskip

\bigskip

\bigskip

\bigskip

\bigskip

\bigskip

\bigskip

\bigskip

\bigskip

\bigskip

\bigskip

{\normalsize \bigskip }

\bigskip

\bigskip

\bigskip

\bigskip

\bigskip

\bigskip

\bigskip

\bigskip

\bigskip

\bigskip

\bigskip

\bigskip

\bigskip

\bigskip

\bigskip

\bigskip

\bigskip

\bigskip

\bigskip

\bigskip

\bigskip

\bigskip

\bigskip

\bigskip

\bigskip

\bigskip

\bigskip

\bigskip

\bigskip

\bigskip

\bigskip

\bigskip

\bigskip

\bigskip

\bigskip

\bigskip

\bigskip

\bigskip

{\Large It is well known that the introduction of a set of auxiliary%
\footnote{{\Large {\large {\normalsize By definition an auxiliary field does
not describe an independent degree of freedom; its equation of motion is
algebraic}.}}} fields is an appropriate framework to the quantization of
theories with on-shell algebra, whereupon the algebra becomes off-shell, see 
}$\left[ 1\right] ${\Large \ and references therein. An example of theories
with on-shell algebra are the BF models }$\left[ 2\right] ${\Large \ and
supersymmetric (susy) theories }$\left[ 3,4\right] ${\Large . Furthermore,
the method of introducing auxiliary fields serves a BRST invariant extension
of the classical action }$\left[ 1,5,6\right] ${\Large . }

{\Large In Ref.}$\left[ 7\right] ${\Large \ ( see also Ref.}$\left[ 8\right] 
${\Large ), it is shown that the introduction of auxiliary fields in the
quantized topological BF theories guarantees also the off-shell exactness of
the full quantum action with respect to the BRST operator. This reports that
the BF theories with auxiliary fields can be formulated from susy theories
of Wess-Zumino type, by the intermediary of the twisting process. This
connection was obtained in dimension two }$\left[ 9\right] ${\Large \ and
four }$\left[ 10\right] ${\Large .}

{\Large However, the exactness of the full N=2, D=4 susy Yang-Mills (SYM)
action is established in Ref.}$\left[ 11\right] ${\Large . The followed
strategy is to use the extension of BRST formalism\footnote{{\Large {\large 
{\normalsize Also called BV or field-antifield formalism [26,27].}}}} to
include the global susy. This suggests a natural variable redefinition
between the N=2, D=4 SYM theories and topological Yang-Mills (TYM) models,
without using twisting procedure }$\left[ 12\right] ${\Large . }

{\Large Furthermore, in Ref.}$\left[ 13\right] ${\Large \ ( see also Ref.}$%
\left[ 14\right] ${\Large \ ) an off-shell closure for N=1, D=4 SYM theories
was constructed functionally by a linearized Slavnov-Taylor operator. This
analysis is also applied for N=2, D=4 SYM theories }$\left[ 15,16\right] $%
{\Large . In this representation the auxiliary fields formalism is available 
}$\left[ 17\right] ${\Large , while the on-shell closure ( with auxiliary
fields eliminated ) of the algebra can be rectified by adding some quadratic
terms in the action, after using the formalism of Batalin-Vilkovisky (BV) }$%
\left[ 18\right] ${\Large . }

{\Large Recently, the quantization of N=1, D=4 SYM theories, without
auxiliary fields, is accomplished by generalized spinor superfields }$\left[
19\right] ${\Large . In this superfield formulation, the action is written
in terms of spinor superfields in the so-called transverse gauge. This
treatment is also considered for N=4 SYM theories }$\left[ 20\right] $%
{\Large .}

{\Large On the other hand, the possibility of directly twisting the N=1 SYM
theory in a "microscopic" topological model was studied in }$\left[ 21,22,23%
\right] ${\Large . }

{\Large In this present paper, we construct a BRST exact extension of the
classical action for N=1, D=4 SYM theories via the introduction of an
auxiliary field, in analogy to what is realized for the case of BF theory }$%
\left[ 7\right] ${\Large . Thus we end up with a BRST exact full quantum
action which leads after the elimination of the auxiliary field, to the same
quantum action obtained in the context of BV formalism }$\left[ 24\right] $%
{\Large \ as well as in the framework of the superfibre bundle approach }$%
\left[ 25\right] ${\Large .}

{\Large In the beginning we will introduce, besides the fields present in
the quantized N=1, D=4 SYM theories, an extrafield which allows an algebraic
construction of an off-shell nilpotent BRST operator. From this, the gauge
fixing action can be written in BRST exact form. Furthermore, we show that
the classical action becomes BRST exact. Therefore, we will see that the
obtained BRST exact complete quantum action makes us observe that the
extrafield\ is a non dynamical auxiliary field. The elimination of this
auxiliary field permits us to recover the standard quantum action with
on-shell nilpotent BRST symmetry.}

\QTP{Body Math}
{\Large First let us recall that the N=1, D=4 SYM theory is build from a
gauge potential }${\Large A}_{\mu }${\Large \ and a four component spinor (
the susy partner of }$A_{\mu }${\Large ) }$\lambda ^{a}${\Large . Both of
them taking values in a Lie algebra of a gauge group }$G${\Large . The
classical action reads\footnote{{\Large {\large {\normalsize All notations
and conventions are as in Wess and Bagger [4].}}}}\ }$\left[ 4\right] $%
\begin{equation}
S_{0}=-\frac{1}{4}F_{\mu \nu }F^{\mu \nu }-\frac{1}{2}\lambda ^{a}\left(
\gamma ^{\mu }\right) _{ab}D_{\mu }\lambda ^{b},  \label{eqn1}
\end{equation}%
{\Large where }$F_{\mu \nu }=\partial _{\mu }A_{\nu }-\partial _{\nu }A_{\mu
}+\left[ A_{\mu },A_{\nu }\right] ${\Large ,\ }$D_{\mu }=\partial _{\mu }+%
\left[ A_{\mu },.\right] ${\Large \ and }$\gamma ^{\mu }${\Large \ the Dirac
matrices in the Weyl basis. In }$\left( 1\right) ${\Large \ and in what
follows the integration sign over the D=4 space-time and trace over the
indices of }$G${\Large \ are omitted for simplicity.}

\QTP{Body Math}
{\Large The classical action would be invariant under susy\ transformations
with parameter }$\omega ${\Large \ given by}%
\begin{equation}
\omega =\theta ^{i}I_{i}+\omega ^{a}Q_{a}+\omega ^{\mu }P_{\mu },
\label{eqn2}
\end{equation}%
{\Large where }$\left\{ I_{i}\right\} _{i=1,..,\dim G}${\Large \ and }$%
\left\{ Q_{a},P_{\mu }\right\} _{a=1,..,4}${\Large \ are the generators of
gauge group and N=1 susy group, respectively. They satisfy the following
commutation relations}%
\[
\left[ I_{i},I_{j}\right] =f_{ij}^{k}I_{k}, 
\]%
\[
\left[ I_{i},P_{\mu }\right] =\left[ Q_{a},P_{\mu }\right] =\left[ P_{\mu
},P_{\nu }\right] =0, 
\]%
\[
\left[ Q_{a},Q_{b}\right] =2\left( \gamma ^{\mu }\right) _{ab}P_{\mu }, 
\]%
\begin{equation}
\left[ I_{i},Q_{a}\right] =b_{i}^{\ast }Q_{a}.  \label{eqn3}
\end{equation}%
{\Large where }$b_{i}^{\ast }=b_{i}${\Large \ for }$a=1,2${\Large \ and }$%
b_{i}^{\ast }=-b_{i}${\Large \ for }$a=3,4${\Large , giving the
representation of the gauge symmetry of }$Q_{a}${\Large .}

{\Large The fields }$\theta ^{i}${\Large , }$\omega ^{a}${\Large \ and }$%
\omega ^{\mu }${\Large \ are the parameters of the Yang-Mills (YM) symmetry,
susy and translation, respectively. From this, we introduce the ghost fields
required for the process of quantization, by substituting, as usual, the
parameters }$\theta ^{i}${\Large , }$\omega ^{a}${\Large \ and }$\omega
^{\mu }${\Large \ by a product of constant anticommuting parameter and the
ghosts }$c^{i}${\Large , }$\chi ^{a}${\Large \ and }$\xi ^{\mu }${\Large ,
respectively. We also need for the gauge-fixing process the antighost }$%
\overline{c}^{i}${\Large , }$\overline{\chi }^{a}${\Large \ and }$\overline{%
\xi }^{\mu }${\Large \ of }$c^{i}${\Large , }$\chi ^{a}${\Large \ and }$\xi
^{\mu }${\Large , respectively, with their corresponding Stueckelberg fields 
}$B${\Large , }$G^{a}${\Large \ and }$E^{\mu }$. {\Large The action of N=1
susy generators }$\left\{ Q_{a},P_{\mu }\right\} ${\Large \ on the fields of
the theory is given by }$\left[ 4\right] $%
\[
\left[ P_{\mu },X\right] =\partial _{\mu }X\text{ \ ,\ \ \ \ \ \ \ \ \ \ \ }X%
\text{ \ \ }{\Large any}\text{ }{\Large field} 
\]%
\[
\left[ Q_{a},A_{\mu }^{i}\right] =-\left( \gamma _{\mu }\right) _{ab}\lambda
^{bi}, 
\]%
\[
\left[ Q_{a},\lambda _{b}^{i}\right] =-\frac{1}{2}\left( \gamma ^{\mu \nu
}\right) _{ab}F_{\mu \nu }^{i}, 
\]%
\[
\left[ Q_{a},c^{i}\right] =\left( \gamma ^{\mu }\right) _{ab}\chi ^{b}A_{\mu
}^{i}, 
\]%
\begin{equation}
\left[ Q_{a},F_{\mu \nu }^{i}\right] =\left( \gamma _{\mu }\right)
_{ab}\left( D_{\nu }\lambda ^{b}\right) ^{i}-\left( \gamma _{\nu }\right)
_{ab}\left( D_{\mu }\lambda ^{b}\right) ^{i}  \label{eqn4}
\end{equation}%
{\Large where} $\gamma ^{\mu \nu }=\frac{1}{2}\left( \gamma ^{\mu }\gamma
^{\nu }-\gamma ^{\nu }\gamma ^{\mu }\right) {\Large .}$

\bigskip

{\Large It is worthwhile to mention that we are interested in our present
investigation on the global N=1 susy transformation. So that, the parameter
of the susy \ and translation groups must be space-time constant, i.e.}%
\[
\]%
\[
\partial _{\mu }\chi ^{a}=0,
\]%
\begin{equation}
\partial _{\mu }\xi ^{\nu }=0,  \label{eqn5}
\end{equation}%
{\Large and satisfy the condition of normalizations}%
\begin{equation}
\overline{\chi }^{a}\chi _{b}=\delta _{b}^{a},  \label{eq6}
\end{equation}%
\begin{equation}
\overline{\xi }^{\mu }\xi _{\nu }=\delta _{\nu }^{\mu }.  \label{eq7}
\end{equation}%
{\Large Thus, in view of }$\left( 5\right) ${\Large \ the classical symmetry
reads}%
\[
\delta A_{\mu }=D_{\mu }\omega =D_{\mu }\theta -\omega ^{a}\left[
Q_{a},A_{\mu }\right] -\omega ^{\nu }\left[ P_{\nu },A_{\mu }\right] ,
\]%
\begin{equation}
\delta \lambda ^{a}=-\left[ \omega ,\lambda ^{a}\right] =-\left[ \theta
,\lambda ^{a}\right] -\omega ^{b}\left[ Q_{b},\lambda ^{a}\right] -\omega
^{\mu }\left[ P_{\mu },\lambda ^{a}\right] .  \label{eq8}
\end{equation}%
{\Large Inserting eqs.}$\left( 4\right) ${\Large \ into }$\left( 8\right) $%
{\Large ,\ we obtain the following on-shell susy transformations }%
\[
\delta A_{\mu }=D_{\mu }\theta +\omega ^{a}\left( \gamma _{\mu }\right)
_{ab}\lambda ^{b}-\omega ^{\nu }\partial _{\nu }A_{\mu },
\]%
\begin{equation}
\delta \lambda ^{a}=-\left[ \theta ,\lambda ^{a}\right] +\frac{1}{2}\omega
^{b}\left( \gamma ^{\mu \nu }\right) _{b}^{a}F_{\mu \nu }-\omega ^{\mu
}\partial _{\mu }\lambda ^{a}.  \label{eq9}
\end{equation}%
{\Large However, an elegant way to overcome the problem of the above
mentioned on-shell closure of the algebra is to use the BV formalism }$\left[
26\right] ${\Large \ ( see also Ref. }$\left[ 27\right] ${\Large ). Its
characteristic feature is the doubling of the field content of the theory by
the introduction of the so-called antifields. These are finally eliminated
by the gauge fixing procedure, leading to the quantum theory in which
effective BRST transformations are nilpotent on-shell }$\left[ 24\right] $%
{\Large . }

{\Large Let us note that the symmetry }$\left( 9\right) ${\Large \ would,
naturally, lead to an on-shell BRST algebra in the minimal sector. Indeed,
we have the following BRST transformations}%
\begin{equation}
Q\lambda ^{ai}=-\left[ c,\lambda ^{a}\right] ^{i}+\frac{1}{2}\chi ^{b}\left(
\gamma ^{\mu \nu }\right) _{b}^{a}F_{\mu \nu }^{i}-\xi ^{\mu }\partial _{\mu
}\lambda ^{ai},  \label{eq10}
\end{equation}%
{\Large \ \ \ }%
\[
QA_{\mu }^{i}=\left( D_{\mu }c\right) ^{i}+\chi ^{a}\left( \gamma _{\mu
}\right) _{ab}\lambda ^{bi}-\xi ^{\nu }\partial _{\nu }A_{\mu }^{i}, 
\]%
\[
Qc^{i}=-\frac{1}{2}\left[ c,c\right] ^{i}+\chi ^{a}\left( \gamma ^{\mu
}\right) _{ab}\chi ^{b}A_{\mu }^{i}-\xi ^{\nu }\partial _{\nu }c^{i}, 
\]%
\[
Q\chi ^{a}=0, 
\]%
\[
Q\xi ^{\mu }=-\chi ^{a}\left( \gamma ^{\mu }\right) _{ab}\chi ^{b}, 
\]%
\[
QF_{\mu \nu }^{i}=-\left[ c,F_{\mu \nu }\right] ^{i}-\chi ^{a}\left[ \left(
\gamma _{\mu }\right) _{ab}\left( D_{\nu }\lambda \right) ^{bi}-\left(
\gamma _{\nu }\right) _{ab}\left( D_{\mu }\lambda \right) ^{bi}\right] -\xi
^{\rho }\partial _{\rho }F_{\mu \nu }^{i}, 
\]%
\[
Q\overline{c}^{i}=B^{i}\text{ \ \ \ \ \ , \ \ \ \ }QB^{i}=0, 
\]%
\[
Q\overline{\chi }^{a}=G^{a}\text{ \ \ \ \ \ , \ \ \ \ \ }QG^{a}=0, 
\]%
\begin{equation}
Q\overline{\xi }^{\mu }=E^{\mu }\text{ \ \ \ \ \ , \ \ \ \ \ }QE^{\mu }=0,
\label{eq11}
\end{equation}%
{\Large where the BRST operator is nilpotent only on-shell, since}%
\begin{equation}
Q^{2}\lambda =\chi \gamma ^{\mu }\chi D_{\mu }\lambda .  \label{eq12}
\end{equation}%
{\Large The right-hand side of }$\left( 12\right) ${\Large \ is proportional
to the equation of motion }$\frac{\delta S_{0}}{\delta \lambda }${\Large \
of }${\Large \lambda }${\Large \ and thus a natural way to overcome this
problem is to modify the BRST transformation }$\left( 10\right) ${\Large \
by introducing an extrafield }$D^{i}${\Large \ \ guaranteeing the off-shell
nilpotency of the BRST symmetry. This is simply realized by adding }$\chi
^{a}D^{i}${\Large \ to the right-hand side of }$\left( 10\right) ${\Large \
and imposing the off-shell nilpotency. So, we get}%
\[
Q\lambda ^{ai}=-\left[ c,\lambda ^{a}\right] ^{i}+\frac{1}{2}\chi ^{b}\left(
\gamma ^{\mu \nu }\right) _{b}^{a}F_{\mu \nu }^{i}-\xi ^{\mu }\partial _{\mu
}\lambda ^{ai}+\chi ^{a}D^{i}, 
\]%
\begin{equation}
QD^{i}=-\left[ c,D\right] ^{i}-\chi ^{a}\left( \gamma ^{\mu }\right)
_{ab}\left( D_{\mu }\lambda ^{b}\right) ^{i}-\xi ^{\mu }\partial _{\mu }D^{i}
\label{eq13}
\end{equation}%
{\Large It is easy to check the off-shell nilpotency of }$\left( 13\right) $%
{\Large \ by an explicit calculation.}

\bigskip

{\Large The next step is to give the complete quantum action of N=1, D=4 SYM
theories and to show how this can be written in BRST exact form. To this
purpose, let us recall that the gauge fixing action obtained in the case of
YM theories is given by}%
\begin{equation}
S_{gf}^{YM}=Q\left( \overline{c}\partial ^{\mu }A_{\mu }\right) ,
\label{eq14}
\end{equation}%
{\Large which involves a Lorentz gauge given by}%
\begin{equation}
\partial ^{\mu }A_{\mu }=0,  \label{eq15}
\end{equation}%
{\Large with the help of the pair }$\left( \overline{c},B\right) ${\Large .}

{\Large In the case of SYM theory we shall choose a susy gauge fixing which
is the extension of the Lorentz gauge\footnote{{\Large {\large {\normalsize %
Such extension is also considered in Ref. [28], where }}$\partial ^{\mu
}A_{\mu }${\normalsize \ \ r}{\large {\normalsize epresents a component of a
chiral superfield.}}}}. This gauge fixing can be obtained from }$\left(
15\right) ${\Large \ by using the following substitution}%
\[
A_{\mu }\longrightarrow A_{\mu }+\left[ \partial _{\mu }\lambda ^{a},Q_{a}%
\right] 
\]%
{\Large with the help of the pair }$\left( \overline{c}+\overline{\chi }%
^{a}Q_{a}+\overline{\xi }^{\mu }P_{\mu },B+G^{a}Q_{a}+E^{\mu }P_{\mu
}\right) ${\Large .}

{\Large Thus, the gauge fixing action }$\left( 14\right) ${\Large \ becomes%
\footnote{{\Large {\large {\normalsize The last term in (16) which involves
the ghosts and antighosts is the susy extension of the usual Faddeev-Popov
action.}}}}}%
\begin{equation}
S_{gf}^{SYM}=Q\left( \overline{c}\partial ^{\mu }A_{\mu }+2b_{i}^{\ast }%
\overline{\chi }^{a}\left( \gamma ^{\mu }\right) _{ab}\partial _{\mu
}\square \lambda ^{bi}\right) .  \label{eq16}
\end{equation}%
{\Large To find this, we have used the following rules}%
\[
Tr\left( I^{i}I_{j}\right) =\delta _{j}^{i}, 
\]%
\[
Tr\left( \left[ Q_{a},Q_{b}\right] \right) =2\left( \gamma ^{\mu }\right)
_{ab}\partial _{\mu }, 
\]%
\begin{equation}
Tr\left( P^{2}\right) =0.  \label{eq17}
\end{equation}%
{\Large On the other hand, a simple calculation with the help of the BRST
transformations }$\left( 11\right) ${\Large \ and }$\left( 13\right) $%
{\Large \ leads to}%
\begin{equation}
QS_{0}=-\chi ^{a}D_{i}\left( \gamma ^{\mu }\right) _{ab}D_{\mu }\lambda
^{bi}.  \label{eq18}
\end{equation}%
{\Large Thus the classical action is not BRST invariant and since our goal
is to construct a BRST exact complete quantum action, we shall construct a
BRST invariant extension of the classical action, which is BRST exact. }

{\Large Employing the BRST transformations }$\left( 11\right) ${\Large , }$%
\left( 13\right) ${\Large \ and the condition }$\left( 6\right) ${\Large ,
we observe that the action }%
\begin{equation}
S_{ex}=Q\left( \frac{1}{2}\overline{\chi }^{a}D^{i}\lambda _{ai}+\frac{1}{16}%
F_{\mu \nu }^{i}\overline{\chi }^{a}\left( \gamma ^{\mu \nu }\right)
_{ab}\lambda _{i}^{b}\right) ,  \label{eq19}
\end{equation}

{\Large can be put in the form}%
\begin{equation}
S_{ex}=S_{0}-\frac{1}{2}D^{i}D_{i}+G^{a}\left( \frac{1}{2}D^{i}\lambda _{ai}+%
\frac{1}{16}F_{\mu \nu }^{i}\left( \gamma ^{\mu \nu }\right) _{ab}\lambda
_{i}^{b}\right) .  \label{eq20}
\end{equation}%
{\Large This represents the effective BRST invariant extension of the
classical action since the last term vanishes as equation of motion of the
Stueckelberg auxiliary field }$G^{a}${\Large .}

{\Large Having found the BRST invariant extension action we now write the
following complete quantum action}%
\begin{equation}
S_{q}=S_{ex}+S_{gf}^{SYM},  \label{eq21}
\end{equation}%
{\Large which is consequently BRST exact. This property provides the
possibility of directly twisting the N=1, D=4 SYM in a topological model.}

{\Large It is worthnoting that the quantum action }$\left( 21\right) $%
{\Large \ permits us to observe that the auxiliary field }$D^{i}${\Large \
is effectively non dynamical, as its equation of motion is a constraint}%
\begin{equation}
\frac{\delta S_{q}}{\delta D^{i}}=-D_{i}+2b_{i}^{\ast }\left( \partial _{\mu
}\square \overline{\chi }^{a}\right) \left( \gamma ^{\mu }\right) _{ab}\chi
^{b}=0.  \label{eqn22}
\end{equation}%
{\Large Thus the essential role of }$D^{i}${\Large \ is to guarantee the
off-shell nilpotency of the BRST operator as well as to establish the
exactness of the quantum action. }

{\Large \bigskip }

{\Large In conclusion, working with the same spirit as in BF theories }$%
\left[ 7\right] ${\Large , we have realized the construction of the
off-shell BRST symmetry for N=1, D=4 SYM theories by introducing, besides
the field content of the theory, an extrafield. Another advantage of our
approach is that the method of introducing auxiliary fields permits us to
perform\ as usual the construction of the BRST exact gauge fixing action, as
well as the BRST exact extension of the classical action. Therefore, we have
build the BRST exact quantum action, which allows us to observe that the
extrafield is a non dynamical auxiliary field. The elimination of this
auxiliary field by means of its equation of motion permits us to recover the
same quantum theory obtained in the context of the BV quantization method }$%
\left[ 24\right] ${\Large \ as well as in the superfibre bundle approach }$%
\left[ 25\right] ${\Large .}

{\Large Finally, we can go a step further, and we can use the method of
introducing auxiliary fields for establishing the exactness of the quantum
action for N\TEXTsymbol{>}1 \ SYM theory, since the general algebra with N%
\TEXTsymbol{>}1 which are called extended susy algebra, can be decomposed
into N=1 sectors }$\left[ 4,28\right] ${\Large . Therefore, the SYM theories
are transformed into topological theories of Witten type by simply
introducing extrafields playing the role of auxiliary fields.}

{\Large \bigskip }

\bigskip

\bigskip

\bigskip

\bigskip \bigskip

\bigskip

\bigskip

\bigskip

\bigskip

\bigskip

\bigskip

\bigskip

\begin{quote}
\bigskip

\bigskip

\bigskip

\bigskip

\bigskip

\bigskip

\bigskip

\bigskip

\bigskip

\bigskip

\bigskip

\bigskip
\end{quote}

\end{document}